\documentclass[letter]{aa}  

\usepackage{xcolor} 
\usepackage{graphicx}
\usepackage{lscape}
\usepackage{multirow}
\usepackage{txfonts}
\bibpunct{(}{)}{;}{a}{}{,} 

\begin{document}

   \title{Chemical segregation in the young protostars Barnard 1b-N and S}

   \subtitle{Evidence of pseudo-disk rotation in Barnard 1b-S\thanks{Based on observations carried out with the IRAM  Northern Extended Millimeter Array (NOEMA). IRAM is supported by INSU/CNRS (France), MPG (Germany), and IGN (Spain).}}

   \author{ A. Fuente \inst{1}
\and M. Gerin  \inst{2}
\and J. Pety \inst{2,3}
\and B. Commer\c con \inst{4}
\and M. Ag\'undez \inst{5}
\and J. Cernicharo \inst{5}
\and N. Marcelino \inst{5}
\and E. Roueff \inst{2}
\and D.C. Lis \inst{2}
\and H.A. Wootten \inst{6}
 }

   \institute{Observatorio Astron\'omico Nacional (OAN,IGN), Apdo 112, E-28803 
Alcal\'a de Henares, Spain.
 \email{a.fuente@oan.es} %
\and
 LERMA, Observatoire de Paris, PSL Research University, CNRS, 
 Sorbonne Universit\'es, UPMC Univ. Paris 06, Ecole Normale Sup\'erieure, F-75005 Paris, France. 
 \and
  Institut de Radioastronomie Millim\'etrique (IRAM),
  300 rue de la Piscine, 38406 Saint Martin d'H\`eres, France.
\and
 Centre de Recherche Astronomique de Lyon (CRAL), Ecole Normale
  Sup\'erieure de   Lyon, CNRS-UMR5574, France.
\and
Instituto de Ciencia de Materiales de Madrid (ICMM-CSIC). E-28049,
  Cantoblanco,  Madrid, Spain.
  \and
National Radio Astronomy Observatory, 520 Edgemont Road, Charlottesville, VA 22903, USA   
}

   \date{Received xxx; accepted xxx}

 
  \abstract
{The extremely young Class 0 object B1b-S and the first hydrostatic
core (FSHC) candidate, B1b-N, provide a unique opportunity to study the chemical changes produced
in the elusive transition from the prestellar core to the protostellar phase. We present 40$"$$\times$70$"$ images of 
Barnard 1b in the $^{13}$CO 1$\rightarrow$0, 
   C$^{18}$O 1$\rightarrow$0, NH$_2$D  1$_{1,1}a$$\rightarrow$1$_{0,1}s,$ and SO 3$_2$$\rightarrow$2$_1$ lines
   obtained with the NOEMA interferometer. The observed chemical segregation allows us to
   unveil the physical structure of this young protostellar system down 
   to scales of $\sim$500~au.    
The two protostellar objects are embedded in an elongated 
condensation, with a velocity gradient of $\sim$0.2-0.4 m s$^{-1}$ au$^{-1}$ in the east-west direction, reminiscent of
an axial collapse. The NH$_2$D data reveal cold 
and dense pseudo-disks (R$\sim$500-1000 au) around each protostar. Moreover, 
we observe evidence of pseudo-disk rotation around B1b-S.
We do not see any signature of the bipolar outflows associated with B1b-N and B1b-S, which were previously 
detected in H$_2$CO and CH$_3$OH, in any of the imaged species.
The non-detection of SO constrains the SO/CH$_3$OH abundance ratio in the high-velocity gas.
}


   \maketitle
%

\section{Introduction}

Barnard 1b (B1b) is a well-known dark cloud
located in the Perseus molecular cloud complex (D=230~pc).
Interferometric observations revealed two young stellar objects (YSOs), B1b-N and B1b-S, 
deeply embedded in a thick envelope \citep{Huang13} that is characterized by its large 
column density, 
N(H$_2$)$\sim$7.6$\times$10$^{22}$~cm$^{-2}$ \citep{Daniel13}, and low kinetic 
temperature, T$_K$ = 12~K \citep{Lis10}.
Based on Herschel fluxes and subsequent spectral energy distribution modeling, \citet{Pezzuto12} 
concluded that B1b-N and  B1b-S were younger than Class 0 sources and proposed 
them to be first hydrostatic core (FHSC) candidates. Using NOEMA, 
\citet{Gerin15} detected and imaged the protostellar ouflows, confirming the young ages
of B1b-N and B1b-S ($\sim$1000 yr and $\sim$3000 yr) and the high-density
environment ($\sim$ a few 10$^5$ cm$^{-3}$ for the outflowing gas). 
ALMA observations of B1b-N and B1b-S, at an unprecedented angular resolution, 
allowed detecting compact condensations of 0.2$"$ and 0.35$"$ radius 
(46 and 80~au) in B1b-N and B1b-S, respectively, in addition
to a more extended envelope \citep{Gerin16}. The sizes and orientations of 
the compact structures are consistent with those of the disks formed in numerical simulations 
of collapsing cores \citep{Gerin16} in the very early phases (before Class 0) of
the star formation process. The young protostars B1b-S and B1b-N are therefore 
promising objects to investigate the beginnings of the low-mass star formation process.

\begin{figure*}[t!]
\hspace{-0.0cm}\vspace{-1.0cm}\includegraphics[width=18cm]{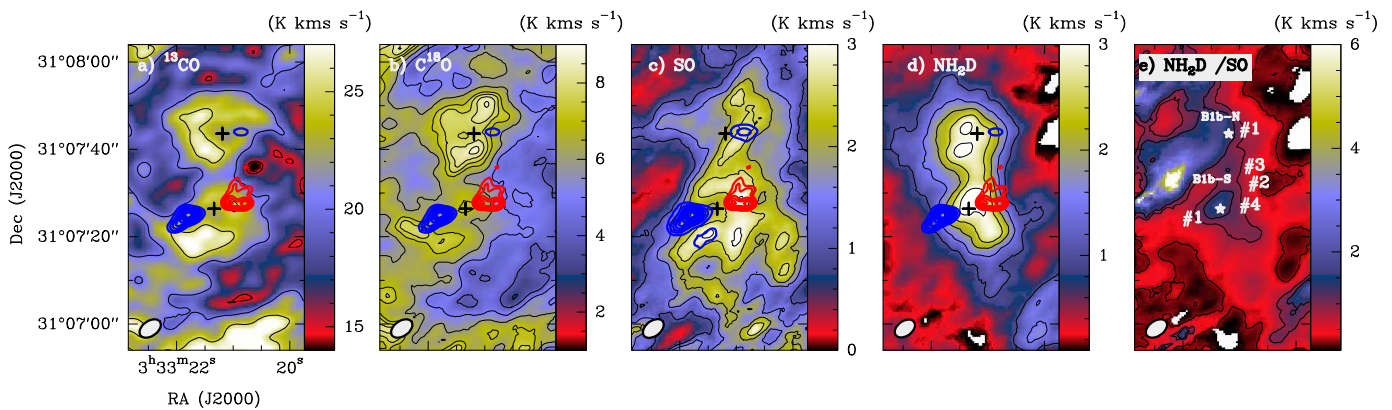}
\caption{{\em a)} Integrated intensity image (NOEMA+30m) of the $^{13}$CO 1$\rightarrow$0 line. Contour levels are 15 
($\approx$150$\times$$\sigma$), 20, and 25 K km s$^{-1}$. Crosses indicate the positions of B1b-N and S. 
The blue and red contours correspond to the blue and red lobes of the B1b-N and S
bipolar outflows as traced by the H$_2$CO emission published by \citet{Gerin15}. 
The beam is plotted in the left bottom corner. 
{\em b)} The same as {\em a)} for the C$^{18}$O 1$\rightarrow$0 line. 
Contour levels are 4 ($\approx$40$\times$$\sigma$), 5, 6, 7, 7.5, 8, and 8.5 km s$^{-1}$.
{\em c)} The same as {\em a)} for the SO 3$_2$$\rightarrow$2$_1$ line. Contour 
levels are 1 ($\approx$10$\times$$\sigma$) to 8 in steps of 0.5 K km s$^{-1}$.
{\em d)} The same as {\em a)} for the integrated intensity of
all the hyperfine components of the NH$_2$D  1$_{1,1}a$$\rightarrow$1$_{0,1}s$ line. Contour levels are 1 
($\approx$10$\times$$\sigma$) to 8 in steps of 0.5 K km s$^{-1}$. 
{\em e)} I(NH$_2$D)/I(SO) integrated intensity ratio (ratio between the images in panels {\em d)} and {\em c)}). 
The spectra toward the positions labeled in this panel are shown in Fig.~\ref{FigA1}. 
}
 \label{Fig1}
\end{figure*}

\section{NOEMA observations}

NOEMA observations were carried out during August and October, 2015 using seven antennas in its CD configurations.
The 3mm receivers were centered on $\sim$110~GHz, covering the 109-111~GHz band.
The wideband correlator WIDEX provided spectroscopic data of the whole band with
a spectral resolution of  $\sim$2~MHz. Several 40~MHz units, providing a spectral resolution of 40~kHz, were 
centered on the frequencies of the intense lines $^{13}$CO 1$\rightarrow$0, C$^{18}$O 1$\rightarrow$0, 
SO  3$_2$$\rightarrow$2$_1$, NH$_2$D 1$_{1,1}a$$\rightarrow$1$_{0,1}s$, and HNCO  5$_{0,5}$$\rightarrow$5$_{0,4}$. 
After 16hr on-source, we obtained an rms of $\sim$0.20~K at 40 kHz spectral resolution 
($\sim$0.1 km s$^{-1}$) 
with an angular resolution of $\sim$$5.0''\times3.4''$ (1155 au$\times$785~au). The HNCO line has 
been barely detected,
and its analysis is not included in this paper.

In order to obtain a high-fidelity image of the entire cloud, we have combined NOEMA observations with short-spacing data. We used the IRAM 30m telescope to map an area twice as large as the requested
NOEMA map, that is, 9 arcmin square, to fully recover the extended structure. 
Short-spacing data were merged with NOEMA observations using the 
\texttt{GILDAS}\footnote{See \texttt{http://www.iram.fr/IRAMFR/GILDAS} 
for  more information about the GILDAS  softwares~\citep{pety05}.}\texttt{/CLASS} software. 
Final images were created  with a velocity resolution of 0.2 km~s$^{-1}$ for SO and the CO 
isotopologues, and of 0.14 km~s$^{-1}$ for the NH$_2$D lines.
In this Letter we use the final 30m+NOEMA cubes (see the observational parameters in Table~\ref{TableA1}).

\section{Large-scale morphology}
The emission of all molecules traces the elongated ridge in which the two protostars and their 
associated outflows are embedded (see the integrated-intensity images in Fig.~\ref{Fig1}). Interestingly, 
the $^{13}$CO 1$\rightarrow$0, C$^{18}$O 1$\rightarrow$0, and  SO 3$_2$$\rightarrow$2$_1$ maps do not 
present emission peaks toward the protostar positions, and only the NH$_2$D image shows local 
peaks toward B1b-N and B1b-S.

The $^{13}$CO 1$\rightarrow$0 brightness temperature
varies between 5 and 11~K along the ridge (see Fig.~\ref{Fig1} and Fig.~\ref{FigA1}). The emission of this transition
is expected to be optically thick in this high-extinction filament \citep{Fuente16}, and its brightness 
temperature is a measure of the gas
kinetic temperature. Assuming that the $^{13}$CO emission is thermalized, which is reasonable for 
densities n(H$_2$)$>$10$^{4}$~cm$^{-3}$, the observed peak brightness temperatures imply gas kinetic
temperatures of T$_k$=8-14~K. These values are in agreement with the averaged temperature of 12~K
obtained from the ammonia lines by \citet{Lis10}. Toward B1b-N, the $^{13}$CO 
and C$^{18}$O 1$\rightarrow$0 lines
present the same intensity, T$_b$$\sim$4~K (Fig.~\ref{FigA1}), consistent with  optically thick gas at T$_k$=8~K.
Toward B1b-S, the line profiles of the $^{13}$CO and C$^{18}$O lines present self-absorption features that
preclude an accurate estimate of the gas temperature. We can only derive a lower limit of T$_k$$>$14~K. 
This limit shows 
that the outflow has already heated the neighborhood of this protostar,
as also demonstrated by 
the high temperatures (30$-$70 K) derived from the CH$_3$OH lines by \citet{Gerin15}.

The SO emission, and to a lesser extent C$^{18}$O, seem to surround the NH$_2$D peaks.
In Fig.~\ref{FigA2} we show the channel velocity maps of the SO and NH$_2$D lines.
The region of enhanced SO emission
clearly surrounds the cold gas traced by NH$_2$D toward B1b-S. Toward B1b-N, the
morphology of the emission is not that clear because the protostellar
core is deeply embedded in the cold filament. It is noticeable that the line width of the SO
line is $\sim$1~km~s$^{-1}$ throughout the region (see Fig.~\ref{FigA1} and Table~\ref{TableA2}). 
The SO emission therefore does not arise from the high-velocity gas associated with the 
bipolar outflows detected by \citet{Gerin15}, but from the dense ridge, as suggested by \citet{Fuente16}. 
 
The morphology of the NH$_2$D emission is most similar to 
that of the 1.2mm continuum emission map reported by \citet{Daniel13}.  
This is confirmed by the  I(NH$_2$D)/I(SO) integrated-intensity map shown in Fig.~\ref{Fig1}e, 
which presents local maxima toward B1b-S and B1b-N.
This deuterated compound is, thus far, our best candidate to probe the neighborhood of the protostars.
In order to explore the gas kinematics in this complex star-forming region, in Fig.~\ref{Fig2}a we show 
the first-order moment map of the main hyperfine 
components of the NH$_2$D 1$_{1,1}a$$\rightarrow$1$_{0,1}s$ line.
The entire molecular filament presents a velocity gradient in the direction perpendicular to
the filament with velocities increasing from  $\sim$6.5 to 7.0 km s$^{-1}$ from
west to east ($\approx$ 0.2$-$0.4 m s$^{-1}$ au$^{-1}$) (see also Fig.~\ref{FigA2}).
This gradient suggests that the whole filament is contracting in the axial direction
and that its fragmentation has driven to the formation of B1b-N and B1b-S. 
This overall east-west velocity gradient twists in the vicinity of B1b-S
because this young protostellar object is formed there. The gas kinematics in the surroundings 
of B1b-S is discussed in detail in Sect.~4. 

\section{B1b-N and S bipolar outflows}
In Fig.~\ref{FigA1} we compare our interferometric spectra with those of the H$_2$CO and 
CH$_3$OH lines published by \citet{Gerin15} toward selected positions in the bipolar outlfow lobes. 
Only the H$_2$CO and CH$_3$OH spectra show high-velocity wings (0.5 to 5.5 km~s$^{-1}$ and 7.5 to 12.5~km~s$^{-1}$
for the B1b-N outflow; $-$4.0 to 5.5 km~s$^{-1}$ and  7.5 to 15 km~s$^{-1}$ in the B1b-S outflow).
The observed C$^{18}$O, SO, and NH$_2$D lines are very narrow with $\Delta$v$<$1~km~s$^{-1}$
(see Table~\ref{TableA2}).
The line width of the $^{13}$CO 1$\rightarrow$0 line is larger, $\Delta$v$\sim$2~km~s$^{-1}$, with emission
in the velocity ranges of 4$-$5.5 km~s$^{-1}$ and 7.5$-$8 km~s$^{-1}$. However,
the channel velocity maps do not show any correlation between the morphology of this moderate velocity gas and 
the lobes of the outflows. More likely, this emission comes from the more turbulent envelope \citep{Daniel13}.

We did not detect high-velocity emission in the SO and $^{13}$CO lines 
with an rms$\sim$0.06~K ($\Delta$v=1~km~s$^{-1}$). 
Assuming optically thin emission and the physical conditions derived by \citet{Gerin15} 
in the outflow positions, T$_k$=20~K and n(H$_2$)=6$\times$10$^{5}$~cm$^{-3}$ in B1b-N and
T$_k$=30~K and n(H$_2$)=5$\times$10$^{5}$~cm$^{-3}$ in B1b-S, we derive
N(SO)/$\Delta$v $<$ 1$\times$10$^{13}$ cm$^{-2}$ km s$^{-1}$ and 
N($^{13}$CO)/$\Delta$v $<$ 2.4$\times$10$^{14}$ cm$^{-2}$ km s$^{-1}$
in the B1b-N outflow; and N(SO)/$\Delta$v $<$1.0$\times$10$^{13}$ cm$^{-2}$ km s$^{-1}$ 
and N($^{13}$CO)/$\Delta$v $<$ 3.0$\times$10$^{14}$ in the outflow associated with B1b-S.
\citet{Gerin15} determined CH$_3$OH column densities 
between N(CH$_3$OH)/$\Delta$v $\sim$ 8$\times$10$^{12}$ cm$^{-2}$ km s$^{-1}$ in B1b-N to
N(CH$_3$OH)/$\Delta$v $\sim$ 1$\times$10$^{14}$ cm$^{-2}$ km s$^{-1}$ in the B1b-S blue lobe.
Even assuming that the high-velocity gas is unresolved by the \citet{Gerin15} observations and 
correcting for the lower angular resolution of our SO image (a factor of 3 in beam areas), 
the N(CH$_3$OH)/$\Delta$v would be a few $\times$ 10$^{13}$ cm$^{-2}$~km~s$^{-1}$
in the B1b-S blue lobe, which is larger than our limit to the SO column density per velocity interval. 
The SO abundance in the B1b-S outflow is therefore lower 
than a few 10$^{-8}$ and hence might be similar to the abundance in the molecular cloud.

\section{Infall signature and pseudo-disk}
 
The line profiles of the $^{13}$CO 1$\rightarrow$0, C$^{18}$O 1$\rightarrow$0 and the main 
component of the NH$_2$D line present self-absorption features at redshifted velocities.
This is consistent with the infall signature expected toward young Class 0 objects.  
The redshifted absorption is not only detected at the protostar positions, but also at some
positions in the outflow lobes (see Fig.~\ref{Fig1} and Fig.~\ref{FigA1}). This suggests that it is a
signature of the global infall of the molecular filament, whose 
contraction has driven the formation of the two protostars. 
One alternative explanation could be that the apparent velocity
gradient is due to the superposition of two filaments with slightly different velocities,
but this interpretation would not explain the high star formation activity in the cloud.
 
Magnetohydrodinamic (MHD) simulations predict that the hydrostatic core is surrounded by a flat and slow rotating structure 
perpendicular to the outflow axis at these earliest
stages of the star
formation process. This structure is commonly referred to as
a pseudo-disk 
(\citealp{Hincelin16}). The detection of the pseudo-disk
remains elusive because the high densities
(n(H$_2$)$>$10$^7$ cm$^{-3}$) and still moderate temperatures of this component
(average kinetic temperature of 10$-$14~K) produce high molecular depletion factors
for most species. Of the species studied in this paper,
only the NH$_2$D emission presents a maximum toward the protostar positions, and
therefore it is our best candidate for the study of the pseudo-disk kinematics.

\begin{figure*}
 \includegraphics[width=1.0\textwidth]{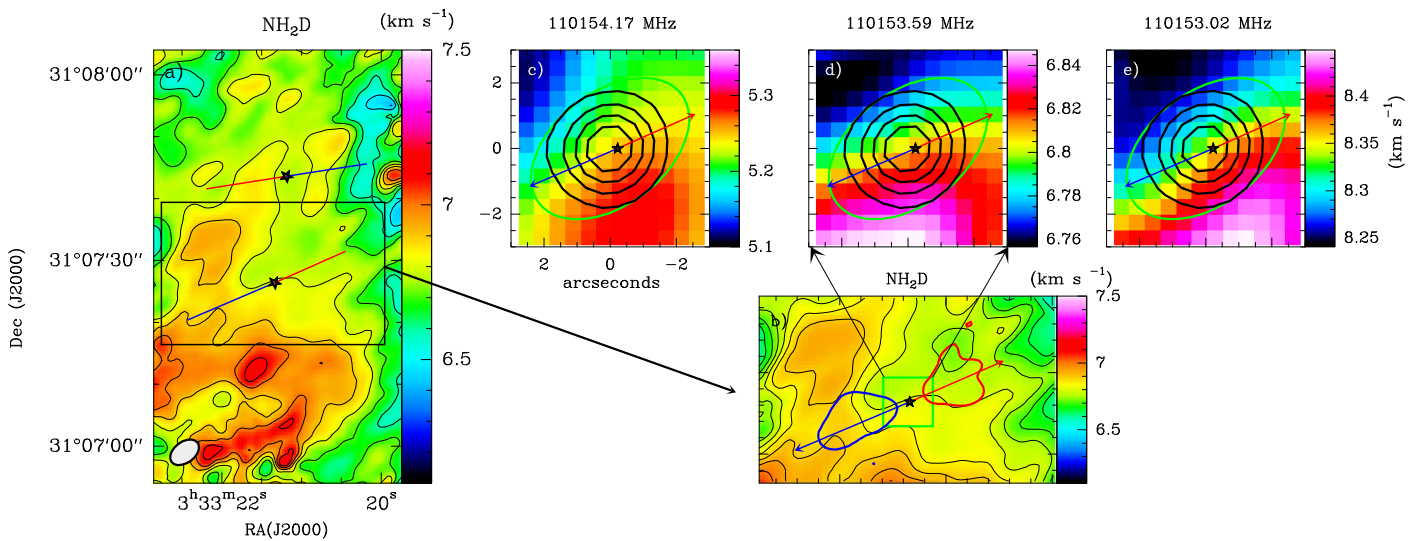}
     \caption{{\em a)} First-order moment map of the hyperfine component of the NH$_2$D line at 110153.59 MHz.
     The stars indicate the positions of B1b-N and B1b-S. Blue and red arrows mark the outflow directions. 
     {\em b)} Zoom around B1b-S of the image shown in {\em a)}. Panels {\em c)}, {\em d),} and {\em d)}
    are the first-order moment maps of the hyperfine components at the of 110154.17, 110153.59, and 110153.03 MHz in 
    a region of 3$"$$\times$3$"$ around B1b-S. The same velocity pattern is observed in the three components, which 
    corroborates
    that the velocity gradient observed in the main component is not an artifact due to self-absorption, but the
    signature of the pseudo-disk rotation. The green ellipse around the star indicates the beam size, and 
    dark black contours correspond to the continuum emission at 145 GHz reported
    by \citet{Gerin15}. Continuum contour levels are 20\% to 90\% in steps of 20\% of the peak emission.}
         \label{Fig2}
   \end{figure*}

In Fig.~\ref{Fig2}b we show the first-order moment map of the main hyperfine 
component of the NH$_2$D 1$_{1,1}a$$\rightarrow$1$_{0,1}s$ line in the surroundings
of B1b-S. The overall east-west velocity gradient that is clearly observed along the whole 
filament is disorted in the vicinity of B1b-S, where the velocity gradient becomes perpendicular to the 
axis of the outflow driven by this protostar. This behavior is also observed in 
the satellite components at 110154.17 and 110153.03 MHz. The detection of the same velocity
pattern in the three hyperfine components confirms that the velocity gradient is real and not an artifact of the self-absorption feature.
We interpret this gradient as the signature of the pseudo-disk rotation predicted by MHD simulations.
According to the rotation axis, the major axis of the pseudo-disk must be located at a PA 
of 45$^{\circ}$ as measured from north to east. This is also consistent with
the elongation of the disk detected with ALMA toward this source \citet{Gerin16}. 
We did not detect any such rotation signature toward B1b-N, which is more deeply embedded.
The angular resolution of our observations is not sufficient to fully resolve the pseudo-disk around B1b-S,
and the detected velocity gradient, 0.2 m~s$^{-1}$ au$^{-1}$, is a lower limit to the rotation 
velocity.  
Toward B1b-S, the  NH$_2$D  1$_{1,1}a$$\rightarrow$1$_{0,1}s$ line presents weak 
wings that are not detected toward adjacent positions. If these wings were caused by 
rotation, the rotation velocity would be $\sim$2.5 km~s$^{-1}$, which is comparable 
to the rotation velocities predicted by the MHD simulations \citep{Galli93,Hincelin16}.

We imaged only one transition per species, and therefore our abundance estimates involve large uncertainties. 
In addition, all the lines may be optically thick in the high-extinction regions around the protostars.
Fortunately, the hyperfine structure of NH$_2$D allows us to estimate the line opacity and 
overcome these limitations.
We fit the NH$_2$D  hyperfine lines assuming the same excitation temperature 
for all the hyperfine components. This procedure allows us to derive
the main line opacities. We obtain opacities in the range of  $\sim$0.24$-$2.0 (see Table~\ref{TableA2}).  
Assuming that the emission uniformly fills the beam, we derive
N(NH$_2$D)$\sim$6.0$\times$10$^{14}$~cm$^{-2}$ and 3.5$\times$10$^{14}$~cm$^{-2}$ toward B1b-N and S,
respectively (Table~\ref{TableA3}).  Using the 1.2mm map reported by \citet{Daniel13}, a constant dust temperature of 12~K and
$\kappa$=0.01~cm$^2$~g$^{-1}$, we estimate N(H$_2$)=1.0$\times$10$^{23}$~cm$^{-2}$ and 1.3$\times$10$^{23}$~cm$^{-2}$
toward B1b-N and S, respectively. Based on these estimates, we derive 
NH$_2$D abundances of  $\sim$6$\times$10$^{-9}$  and $\sim$2$\times$10$^{-9}$  
for B1b-N and S. These values are in agreement with those derived by \citet{Daniel13} for the inner
R$<$4000~au part of the envelope. For SO, we assume that the lines are thermalized at T$_{k}$=10~K and derive 
N(SO)$\sim$1.8$\times$10$^{14}$~cm$^{-2}$ in B1b-N and N(SO)$\sim$1.5$\times$10$^{14}$~cm$^{-2}$ in
B1b-S. The SO abundances 
are $\approx$2$\times$10$^{-9}$ in B1b-N and 1$\times$10$^{-9}$ in B1b-S, a factor of $\sim$10 lower 
than the single-dish estimate by \citet{Fuente16}.

\section{Modeling the gas chemistry}
\begin{figure}[t!]
\hspace{-0.0cm}\includegraphics[width=8cm]{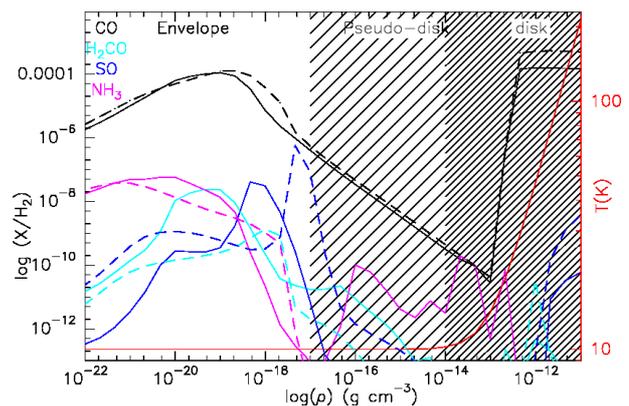}\\

\caption{Chemical model predictions for t=10$^5$ yr assuming as initial abundances the low-metallicity 
case of \citet{Wiebe03} (continuous lines) and solar abundances (dashed lines). We indicate the parameter 
space corresponding to the envelope, pseudo-disk, and 
disk according with the MHD simulations reported by \citet{Gerin16}.
}
 \label{Fig3}
\end{figure}

\citet{Gerin16} compared recent ALMA continuum observations at 349 GHz with the model of
a strongly magnetized, turbulent, collapsing 1 M$_\odot$ core from \citet{Henne16}, with
ambipolar diffusion as in \citet{Masson16}, at an inclination angle of 20$^\circ$. 
The initial core density and temperature were uniform, $\rho_0$=9.4$\times$10$^{-18}$ g cm$^{-3}$
and T=10 K, and the mass-to-flux ratio $\mu$=2. After 6.4$\times$10$^4$ yr, the simulations
show a structure consistent with ALMA and NOEMA observations,
with a disk of R=50$-$75 au ($\rho$$\sim$10$^{-14}$$-$10$^{-11}$ g cm$^{-3}$), a pseudo-disk of
R=700 au ($\rho$$\sim$10$^{-17}$$-$10$^{-14}$ g cm$^{-3}$), and an outflow that extends up
to 1100 au. In this model the gas thermal behavior is described by the barotropic law
$T_{\rm K}  = 10 \times (1+ ( \rho/\rho_c)^{4/3} )^{0.5}$. We used the pseudo-time-dependent  model described by
\citet{Pacheco15,Pacheco16} to calculate the abundances of CO, H$_2$CO, SO, and NH$_3$
at t=10$^5$ yr assuming the physical conditions given by the barotropic law.
The chemical model is an updated  version  of the model reported  
by  \citet{Agundez08} and \citet{Fuente10}, and it includes a complete gas-phase chemical network for C, O, N, and S,
adsorption onto dust grains and desorption processes such as thermal evaporation, photodesorption,
and desorption induced by cosmic rays. The abundances of most molecules have not reached the steady-state
value at these early times, and the results are sensitive to the initial conditions and chemical age. We repeated the calculations
for two sets of initial conditions that are representative of i) the dark cloud case (low-metallicity case of
\citet{Wiebe03}) and ii) solar abundances. In both cases, we adopted $\zeta$=5$\times$10$^{-17}$ s$^{-1}$
as derived by \citet{Fuente16}. As a general behavior, the abundances of the four molecules are
heavily depleted in the pseudo-disk and increase again in the disk where the temperature rises above 100 K (see Fig.~\ref{Fig3}). 
The abundance of H$_2$CO follows that of CO, which starts to freeze out onto the grains in the envelope.   
The SO and NH$_3$ abundances remain relatively high for H$_2$ densities 
below 10$^7$ cm$^{-3}$ ($\rho$=3.8$\times$10$^{-17}$ g cm$^{-3}$). The abundance of 
SO reaches its maximum value at around R=700$-$1000 au, that is, in the outer part of the pseudo-disk.
This model explains the morphology in B1b quite well, where the SO emission surrounds the NH$_2$D cores
while $^{13}$CO and C$^{18}$O are tracing the external envelope, although it fails to account for the observed abundances.
We note that surface chemistry is not included,  and the abundance of some molecules such as NH$_3$ 
could therefore be underestimated. Unfortunately, our model does not include deuterium 
chemistry for further comparison.

\section{Conclusions}
   The combination of NOEMA and 30m data reveals for the first time the physical structure 
   of the young protostellar system
   formed by B1b-N and B1b-S. The two protostellar objects are embedded in an elongated 
condensation with a velocity gradient in the east-west direction
that is reminiscent of
an axial collapse. Our NH$_2$D data allow us to detect the pseudo-disk associated with each protostar. Moreover,
we find evidence of rotation in the pseudo-disk associated with B1b-S.
The envelope, outflows, and pseudo-disks can be chemically segregated:
the envelope is detected in the CO isotopologues; the outflows, in H$_2$CO and CH$_3$OH; and the pseudo-disks,
in NH$_2$D. 
The SO emission seems to come from the 
outer part of the pseudo-disk. It is interesting to note that SO is not detected in the outflows, 
although the sensitivity of our observations does not allow us
to signficantly constrain its abundance
in the high-velocity gas.

\begin{acknowledgements}

We thank the Spanish MINECO for funding support from
grants CSD2009-00038, AYA2012-32032, AYA2016-75066-C2-1/2-P, and ERC under ERC-2013-SyG, G. A. 610256 NANOCOSMOS.
ER and MG thank the INSU/CNRS program PCMI for funding.

\end{acknowledgements}

%

%
%

\begin{appendix}

\section{Additional tables and figures}

\begin{table*}
\caption{Summary of observations}
\label{TableA1}
\begin{tabular}{lc|c|ccc|ccc}
\hline
\hline
\multicolumn{2}{c|}{} & \multicolumn{1}{c|}{} & \multicolumn{3}{c|}{NOEMA}  & \multicolumn{3}{c}{NOEMA+30m} \\ 
Line                       & Freq  (MHz) & E$_{upp}$ (K)  & Beam ($"$) & $PA(^o)$ &  $rms(K)$$^1$ &   Beam ($"$) & $PA(^o)$ &  
$rms(K)$$^1$    \\ \hline
C$^{18}$O 1$\rightarrow$0  &  109782.17  &  5.3    &  $5.01\times3.41$  & $-$52 & 0.13 & $5.48\times3.45$ & $-$54 & 0.12 \\ 
$^{13}$CO 1$\rightarrow$0  &  110201.35  &  5.3       & $4.72\times3.21$  & $-$51 & 0.13 & $5.38\times3.38$ & $-$53 & 0.12 \\ 
NH$_2$D  1$_{1,1}a$$\rightarrow$1$_{0,1}s$ &  110153.59 &   21.3    &$4.74\times3.23$  & $-$51 & 0.16 & $5.21\times3.38$ & $-$54 & 0.14 \\
SO 3$_2$$\rightarrow$2$_1$ &  109252.22  &  21.1  & $4.75\times3.25$  & $-$52 & 0.15 & $5.44\times3.44$ & $-$52 & 0.12 \\
\hline
\hline
\end{tabular}

\noindent
$^1$ rms in a velocity channel of 0.2~km~s$^{-1}$ for C$^{18}$O, $^{13}$CO, and SO, and of 0.14~km~s$^{-1}$ for NH$_2$D.  
\end{table*}

\begin{table*}
\caption{Line fits}
\label{TableA2}
\begin{tabular}{l|l|ccc|ccc|cccc}
\hline
\hline
\multicolumn{2}{c|}{{\small Offsets$^1$}} & \multicolumn{3}{c|}{{\small C$^{18}$O}}  & 
\multicolumn{3}{c|}{{\small SO}} &  \multicolumn{4}{c}{{\small NH$_2$D}} \\ 
\multicolumn{2}{c|}{} & \multicolumn{1}{c}{{\small Area}}  & \multicolumn{1}{c}{{\small $V_{lsr}$}} &  
\multicolumn{1}{c|}{{\small $\Delta V$}} &
\multicolumn{1}{c}{{\small Area}}  & \multicolumn{1}{c}{{\small $V_{lsr}$}} &  \multicolumn{1}{c|}{{\small $\Delta V$}} &
\multicolumn{1}{c}{{\small $(T_{ex}-T_{bg}) \times \tau$}}  & \multicolumn{1}{c}{{\small $V_{lsr}$}} &  
\multicolumn{1}{c}{{\small $\Delta V$}} & \multicolumn{1}{c}{{\small $\tau$}} \\
 \multicolumn{2}{c|}{} & \multicolumn{1}{c}{{\small $(K$ $km$ $s^{-1})$}}  & \multicolumn{1}{c}{{\small $(km$ $s^{-1})$}} &  
 \multicolumn{1}{c|}{{\small $(km$ $s^{-1})$}} &
\multicolumn{1}{c}{{\small $(K$ $km$ $s^{-1})$}}  & \multicolumn{1}{c}{{\small $(km$ $s^{-1})$}} &  
\multicolumn{1}{c|}{{\small $(km$ $s^{-1})$}} &
\multicolumn{1}{c}{{\small $(K)$}}  & \multicolumn{1}{c}{{\small $(km$ $s^{-1})$}} &  \multicolumn{1}{c}{{\small $(km$ $s^{-1})$}} &
 \multicolumn{1}{c}{} \\
 \hline
{\small B1b-N}   &  {\small (0,0)}  & {\small 7.50(0.09)} &   {\small 6.9(0.1)} &  {\small 1.4(0.1)} & {\small 2.02(0.05)} & 
{\small 6.6(0.1)} &  {\small 0.7(0.1)} &  {\small 4.51(0.31)} &  {\small 6.7(0.1)} & {\small 0.9(0.1)} &  {\small 1.7(0.2)} \\ 
{\small N-out\#1} & {\small (-5,1)}  & {\small 6.72(0.08)} &  {\small 6.8(0.1)} &  {\small 1.2(0.1)} & 
{\small 2.49(0.06)} & {\small 6.6(0.1)} &  {\small 0.6(0.1)} &  {\small 1.60(0.17)} &  {\small 6.9(0.1)} &  
{\small 1.3(0.1)} &  {\small 0.4(0.3)} \\ \hline
{\small B1b-S}   &  {\small(0,0)} &  {\small 7.22(0.09)} &  {\small 6.8(0.1)} & {\small 1.6(0.1)} & 
{\small 1.89(0.07)} & {\small 6.9(0.1)} &  {\small 0.9(0.1)} &  {\small 6.18(0.28)} &  {\small 6.8(0.1)} &  {\small 0.7(0.1)} &  
{\small 1.5(0.2)} \\
{\small S-out\#1} &  {\small (6,-2)}  &  {\small 6.54(0.10)} & {\small 6.9(0.1)} & {\small 1.6(0.1)} & {\small 1.94(0.06)} & 
{\small 6.9(0.1)} &  {\small 0.8(0.1)} & {\small 4.20 (0.32)}  & {\small 6.8(0.1)} & {\small 0.6(0.1)}  & {\small 1.5(0.3)} \\
{\small S-out\#2}  & {\small (-9,6)}  &  {\small 4.83(0.08)} & {\small 6.9(0.1)} & {\small 1.2(0.1)} & {\small 2.21(0.06)} & 
{\small 6.6(0.1)} &  {\small 0.7(0.1)} &                     &                  &                    &  
\\   
%
%
{\small S-out\#3}  & {\small (-8,10)} &  {\small 4.79(0.07)} & {\small 6.9(0.1)} &  {\small 1.3(0.1)} & {\small 2.17(0.06)} & 
{\small 6.6(0.1)} &  {\small 0.7(0.1)} &                     &                    &                   &  \\
{\small S-out\#4}  & {\small (-7,1)}  &  {\small 5.52(0.08)} &  {\small 6.9(0.1)} &  {\small 1.4(0.1)} & {\small 2.66(0.07)} & 
{\small 6.7(0.1)} &  {\small 0.9(0.1)} &                     &                    &                    &  
\\ 
\hline
\hline
\end{tabular}

\noindent
{\small $^1$Offsets are given in arcseconds relative to each protostellar object.}

\end{table*}

\begin{table*}
\caption{Molecular column densities}
\label{TableA3}
\begin{tabular}{lcccc}
\hline
\multicolumn{2}{c}{Offsets} & \multicolumn{1}{c}{N(SO)$^a$} &  
\multicolumn{1}{c}{N(NH$_2$D)$^b$}  & \multicolumn{1}{c}{N(NH$_2$D)/N(SO)}
\\ 
 \hline
B1b-N   & (0,0)    &   1.8$\times$10$^{14}$  &    6.0$\times$10$^{14}$  &  $\sim$3.3  \\ 
%
%
B1b-S   & (0,0)    &   1.5$\times$10$^{14}$  &    3.5$\times$10$^{14}$  &  $\sim$2.3    \\
%
%
%
%
\hline \hline
\end{tabular}

\noindent
$^a$Assuming local thermodynamic equilibrium with T$_k$=10 K. \\
\noindent
$^b$ Using the opacities and excitation
temperatures derived from the line fitting (see Table \ref{TableA2}).
\end{table*}

\begin{figure*}[t!]
\hspace{+0.0cm}\includegraphics[width=18cm]{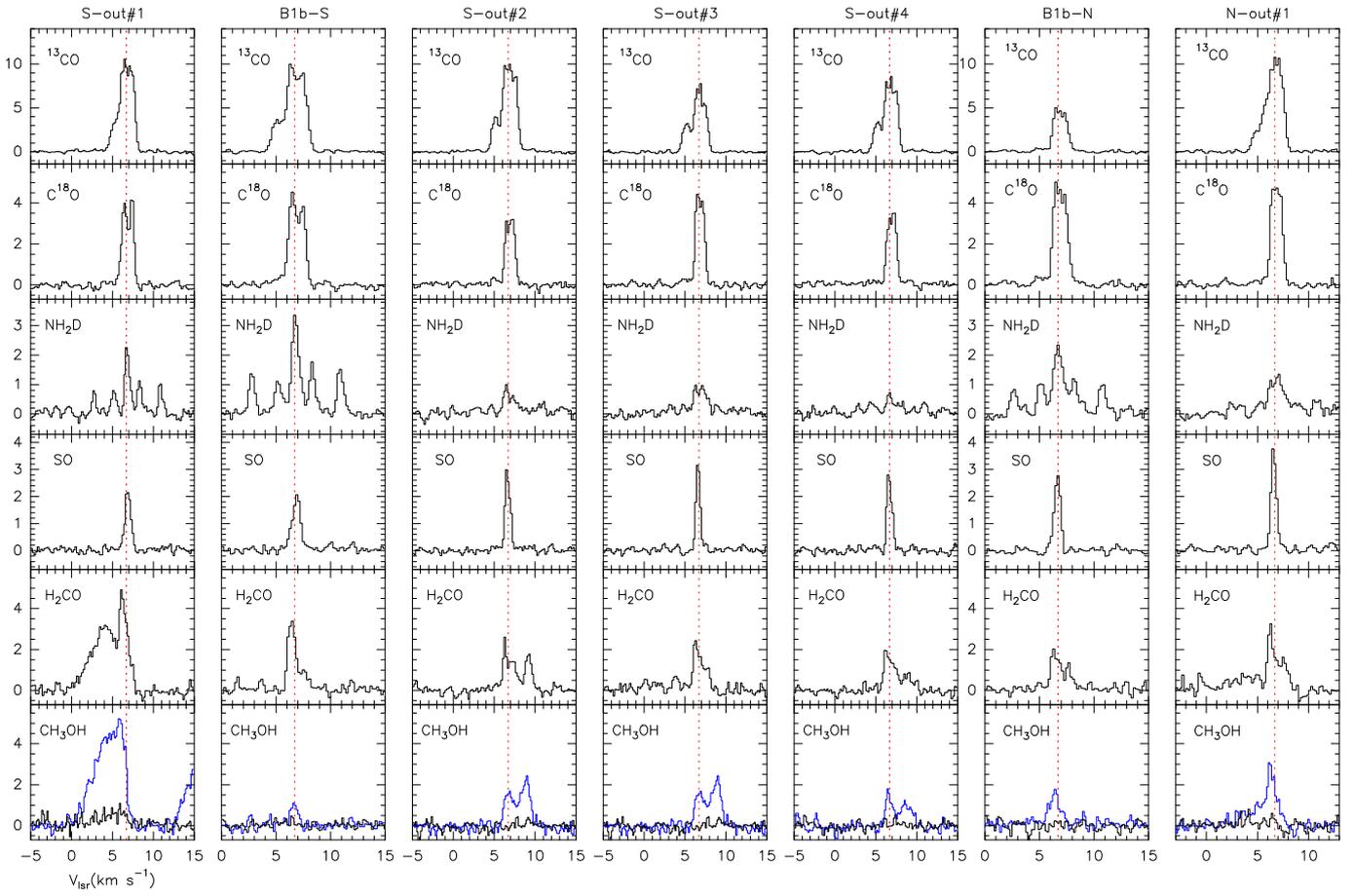}
\caption{Interferometric spectra toward a set of selected positions (intensities in K). In addition to our data, 
we show the spectra of the H$_2$CO 2$_{0,2}$$-$1$_{0,1}$, CH$_3$OH 3$_0$$-$2$_0$ A (blue) and  CH$_3$OH 3$_2$$-$2$_2$ E (black) lines published
by \citet{Gerin15}. The systemic velocity, v$_{lsr}$=6.7~km~s$^{-1}$, is marked with a dashed red line.}
 \label{FigA1}
\end{figure*}


  \begin{figure*}
 \includegraphics[width=0.8\textwidth]{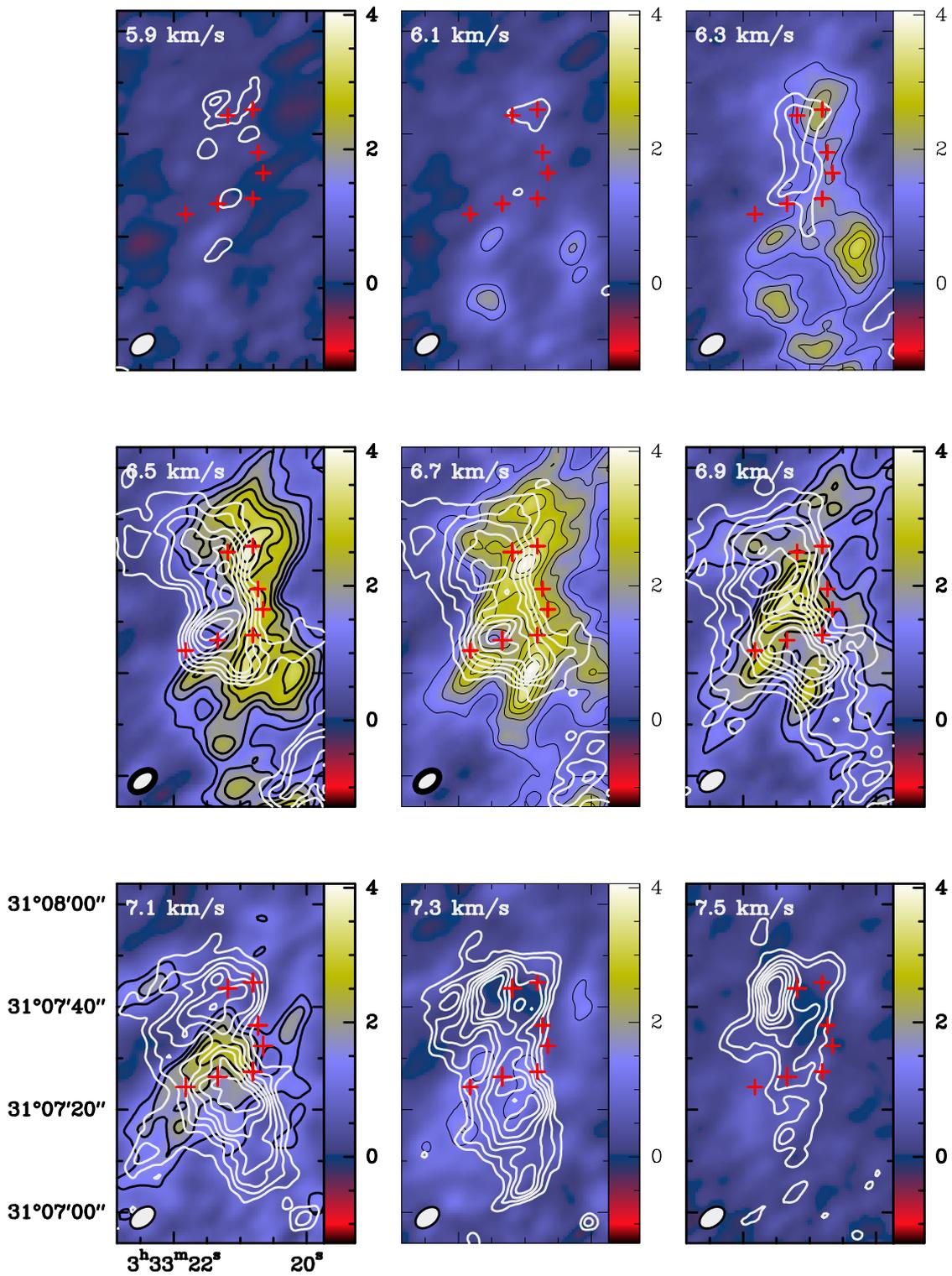}
     \caption{Channel maps of the emission of the SO 3$_2$$\rightarrow$2$_1$ (color map)
and the main component of the NH$_2$D  1$_{1,1}a$$\rightarrow$1$_{0,1}s$ (white contours) lines. The images 
have been created by merging the
NOEMA data with the short spacing obtained with the IRAM 30m telescope. The channel velocity is marked
at the top of the panel. Red crosses indicate the positions studied in detail in this paper.
              }
         \label{FigA2}
   \end{figure*}

\end{appendix}

\end{document}